# Influence of Mg deficiency in MgB$_2$ single crystals on crystal structure and superconducting properties

N. D. Zhigadlo, S. Katrych, J. Karpinski and B. Batlogg

*Laboratory for Solid State Physics, ETH Zurich, 8093 Zurich, Switzerland*

F. Bernardini and S. Massidda

*Department of Physics, University of Cagliari,
Cittadella Universitaria, I-09042 Monserrato (CA), Italy*

R. Puzniak

*Institute of Physics, Polish Academy of Sciences, Aleja Lotników 32/46, 02-668 Warsaw, Poland*

**Abstract.** The effects of high-temperature vacuum-annealing induced Mg deficiency in MgB$_2$ single crystals grown under high pressure were investigated. As the annealing temperature was increased from 800 to 975 °C, the average Mg content in the MgB$_2$ crystals systematically decreased, while $T_c$ remains essentially unchanged and the superconducting transition slightly broadens from ∼ 0.55 K to ∼ 1.3 K. The reduction of the superconducting volume fraction was noticeable already after annealing at 875 °C. Samples annealed at 975 °C are partially decomposed and the Mg site occupancy is decreased to 0.92 from 0.98 in as-grown crystals. Annealing at 1000 °C completely destroys superconductivity. X-ray diffraction analysis revealed that the main final product of decomposition is polycrystalline MgB$_4$ and thus the decomposition reaction of MgB$_2$ can be described as 2MgB$_2$(s) → MgB$_4$(s) + Mg(g). First-principles calculations of the Mg$_{1-x}$(V$_{Mg}$)$_x$B$_2$ electronic structure, within the supercell approach, show a small downshift of the Fermi level. Holes induced by the vacancies go to both σ and π bands. These small modifications are not expected to influence $T_c$, in agreement with observations. The significant reduction of the superconducting volume fraction without noticeable $T_c$ reduction indicates the coexistence, within the same crystal, of superconductive and non-superconductive electronic phases, associated with regions rich and poor in Mg vacancies.

Since the discovery of superconductivity in MgB$_2$ [1], much effort has been devoted to investigate factors which influence its critical temperature, $T_c$. Chemical doping, which modifies the carrier density reduces $T_c$ (see recent review and references therein [2]). It has been suggested that non-stoichiometry may be an important factor that influences superconductivity of MgB$_2$ [3]. Non-stoichiometry is characteristic for many compounds and is often accommodated in the form of vacancies in the crystal. Many metal borides have a very narrow homogeneity range and, for example, the Mg – B phase diagram indicate that MgB$_2$ is a line compound with no solid solution range [4, 5]. The effect of Mg deficiency on structural and superconducting properties has been studied on polycrystalline samples and results are still controversial. For example, studies [6,7] of polycrystalline samples prepared by solid state reaction under magnesium deficiency conditions Mg$_{1-x}$B$_2$ ($0 \leq x \leq 0.5$) revealed the formation of two-phase samples (MgB$_2$ + MgB$_4$) containing small amounts of MgO and other impurity phases. Sharma *et al.* [8] saw $T_c$ depressions of more than 10 K for a nominal composition of Mg$_{0.85}$B$_2$. Ordered Mg vacancies were seen in electron diffraction experiments, with phase separation into Mg vacancy-rich regions and Mg vacancy-poor regions. According to the Rietveld refinement Mg vacancy up to 5 % was reported [9], while Hinks *et al.* [10] found by neutron diffraction experiment only small Mg vacancy concentration (up to 1 %) in their MgB$_2$ compounds with various Mg:B ratios. Further investigations of Mg$_{1-x}$B$_2$ single-phase samples [11-17] showed that the homogeneity region of magnesium diboride is in the range $0 \leq x \leq 0.1$ and the samples with the composition Mg$_{0.9}$B$_2$ [15] are multi-phase and have inhomogeneous microstructure. In the homogeneity region of the Mg$_{1-x}$B$_2$ system $T_c$ slightly decreases by 1 – 3 K with an increase in the concentration of Mg vacancies [11-13, 18]. The range of observations in the literature about the level of non-stoichiometry and its influence on superconducting properties is possibly related to the differences in composition of raw materials, the methods of preparation, impurities etc. Small deviations in stoichiometry cannot be easily measured experimentally especially in polycrystalline samples, which are not highly crystalline and usually contain some MgO as impurity. It is important to notice that in the majority of the published papers the Mg deficiency in $MgB_2$ has not been defined experimentally but just assumed based on an initial (nominal) composition.

The main goal of this work is to provide a clearer picture about the effect of high temperature vacuum annealing on crystal structure and superconducting properties of $MgB_2$ single crystals. The main question being addressed is whether small variations in over-all stoichiometry produce discernible differences in superconducting properties.

High-quality single crystals of $MgB_2$ have been grown under high pressure using the cubic anvil press. The pressure and temperature conditions for the growth of $MgB_2$ single crystals were determined in our earlier study of the Mg – B – N phase diagram [2]. A mixture of magnesium (Fluka, 99.99 % purity), amorphous boron (Alfa Aesar, > 99.99 %), and boron nitride (Saint-Gobain Advanced Ceramics, > 99 %) was placed in a BN crucible. The crucible was surrounded by a graphite heater sleeve and placed in a pyrophyllite cube at the center of the six anvils generating the high pressure. The typical growth process involves: (i) increasing the pressure up to 30 kbar, (ii) increasing the temperature up to 1960 °C in 1 h, (iii) dwelling for 0.5 – 1 h, (iv) lowering the temperature and pressure in 1 h. As a result, $MgB_2$ crystals sticking together with BN crystals have been obtained. It was observed [2, 18] that unsubstituted as-grown $MgB_2$ crystals usually are deficient in Mg by about 2 – 4 %. The Mg deficiency can arise during the crystal growth process and also as a result of the subsequent annealing. Usually, after finishing the crystal growth process under high pressure, the synthesized molten lump is exposed to heating in a quartz ampoule for ~ 0.5 h in dynamic vacuum at 750 °C to evaporate the surplus of magnesium. In order to determine $T_c$, the magnetic moment of an individual crystal was measured in a magnetic field of 5 Oe after cooling in zero field (ZFC) using a commercial Quantum Design superconducting quantum interference device (SQUID) magnetometer. The field was applied parallel to the crystallographic *c*-axis of the sample and $T_c$ is defined as the onset temperature of diamagnetism. For high temperature vacuum annealing MgB$_2$ crystals have been kept for 2 or 6 hours in vacuum at 800, 850, 875, 900, 950, 975, and 1000 °C in a Tempress horizontal tube furnace. After each annealing step, the shape and the surface of the crystals was examined in an optical microscope mounted on X-ray diffractometer to exclude possible mechanical damage of the crystals, affecting on obtained results. Then, the crystals were characterized by single crystal X-ray diffraction and SQUID magnetometry. The crystal structure was investigated at room temperature with an X-ray single-crystal Xcalibur PX, Oxford Diffraction diffractometer equipped with a charge coupled device (CCD) area detector (at the Laboratory of Crystallography, ETH Zurich), which allowed us to examine the whole reciprocal space (Ewald sphere). Data reduction and the analytical absorption correction were introduced using the CRYSALIS software package [19]. The Mg deficiency was estimated using a refinement on $F^2$ [20], by allowing the average Mg occupancy to vary as a free variable. Indeed, the Mg site was found not to be fully occupied.

To perform annealing experiments two MgB$_2$ crystals characterized by almost identical transition temperature of about 38.7 K and importantly with significantly different shapes were selected. They have been placed in a quartz ampoule connected to the vacuum pump and when the pressure was lower than $10^{-6}$ Torr the temperature of the furnace was raised to 800 – 1000 °C at a rate of 300 °C/h, maintained 2 or 6 h and then cooled to room temperature at a rate of 200 °C/h. To exclude possible oxidation resulting from residual oxygen a piece of Ta foil has been placed in the quartz ampoule together with crystals.

Magnetic measurements were performed to trace the changes in temperature dependences of zero field cooled magnetization, being an outcome of appearance of phase separation and of variation in $T_c$. All of the temperature dependences of magnetization $M(T)$ were recorded exactly at the same experimental conditions, i.e., in the same external magnetic field of 5 Oe, applied after zero field cooling exactly parallel to the *c*-axis of the crystal to avoid any impact of the variation of demagnetizing field on the obtained results. The studied crystals were relatively small and the accuracy of their mass determination was not too big and therefore, in order to compare the impact of annealing on their magnetization, for each of distinguishable crystal, obtained $M(T)$ dependences were normalized to ZFC magnetization value at 5 K of the as-grown crystal. Figure 1 presents representative set of magnetic data obtained for one of the studied crystals. The normalized diamagnetic moment $M$ versus temperature is shown for an as-grown $MgB_2$ single crystal and after annealing between 850 and 975 °C. No changes in the onset and width of the superconducting transition ($\Delta T_c$) were found in single crystals annealed below 850 °C. At higher annealing temperatures the diamagnetic signal decreases systematically, indicating a reduction of the superconducting volume fraction of the crystals, while the onset temperature did not change. The appearance of multi-steps in the magnetization curve is not observed, thus, there is no evidence for possible macroscopic phase separation into superconducting regions with significantly different distinct $T_c$. However, most likely, the significant reduction of the superconducting volume fraction without noticeable reduction of $T_c$ is caused by inhomogeneous distribution of the magnesium vacancies in the samples annealed at high temperature, i.e., by selective decrease of Mg site occupancy during annealing, leading to a phase separation into Mg vacancy-poor regions with $T_c$ of as-grown crystals and essentially non-superconducting Mg vacancy-rich regions. A clear effect of high temperature vacuum annealing is observed in the form of a slight broadening of the transition ($\Delta T_c$, 10 – 90 % criterion), (Fig. 1b). The sharp superconducting transition in as-grown crystals ($\Delta T_c \approx 0.55$ K) becomes broader already after annealing at 875 °C, indicating the appearance of structural disorder and reach a value $\Delta T_c \approx 1.3$ K after annealing at 975 °C. Samples annealed at 1000 °C are not superconductors any more. The rapid changes in superconducting volume fraction and $\Delta T_c$ after annealing at the temperatures higher than 875 °C reveal the gradual loss of Mg by vaporization. The observed broadening of the transition region ($\Delta T_c$) reflects a rather small and gradual decrease of $T_c$ upon Mg loss, and crystallographic studies are expected to provide some insight into microscopic mechanism.

Figure 2 depicts typical XRD patterns of annealed $MgB_2$ crystals. No visible change in the diffraction pictures can be observed till 950 °C, which indicates that the $MgB_2$ phase remains stable. However, when crystals were annealed at 975 °C, secondary phases begin to segregate from the $MgB_2$ phase. The phase decomposition becomes more pronounced at higher temperature. These significantly decomposed crystals were not superconducting, completely lost their shiny surface and XRD results confirmed the polycrystalline nature of the samples, with $MgB_4$ as the main phase (Fig. 2 e, f). $MgB_4$ is the product expected when $MgB_2$ decomposes due to Mg loss; according to

$$2MgB_2(s) \rightarrow MgB_4(s) + Mg(g) \qquad (1)$$

The refinement data for $MgB_2$ crystals annealed at various temperatures are presented in Table 1. The Mg deficiency slightly increased with increasing annealing temperature. After annealing at 975 °C, the Mg site occupancy decreased to 0.92 from 0.98 in the as-grown crystals. Even for such heavily affected crystals the onset of superconductivity remains at 38.7 K, indicating that the $MgB_2$ phase is quite robust. The reduction of volume fraction is suggestive of an inhomogeneous distribution of magnesium vacancies. Vacancies can order locally and form a variety of superlattice patterns [8].

Figure 3 shows the lattice parameters *a* and *c* for various Mg vacancy levels. Over-all the lattice parameters are remarkably unaffected by the Mg vacancies, reflecting the stability of the B network. For small deficiency (up to ~ 6 %), the lattice parameter *c* appears to slightly decrease by ~ 0.012 %, while *a* is almost constant. For 8 % of Mg vacancies both parameters slightly increase suggestive of the appearance of new structural defects, possibly associated with the formation (nucleation) of the $MgB_4$ phase, which co-exists in the sample (Fig. 2d). The data in the figure described by open symbols are taken from ref. [21].

Figure 4 shows a summary of our results on Mg occupation and relative diamagnetic signal as a function of annealing temperature. The data indicate that annealing of $MgB_2$ crystals at temperatures up to 850 °C leaves the magnetic signal essentially unaffected. Even after annealing for 6 h at 850 °C crystals do not decompose and we cannot see any polycrystalline rings in the X-ray patterns (Fig. 2b). However, after annealing at 875 °C the relative diamagnetic signal is reduced by almost 20 % which is suggests the formation non-superconducting domains, though the X-ray data are still that of a single crystal. This process is not simple since it involves simultaneously the evaporation of Mg and at temperature $T \geq$ 975 °C the nucleation of polycrystalline $MgB_4$ phase. The process of evaporation of Mg from $MgB_2$ crystals in vacuum at high temperature is distinctly different from the oxidation process of $MgB_2$ in air at high temperature. Stable MgO does not form in vacuum, and the evaporation of Mg is the dominant process.

We compare our observations with already published data and discuss why $T_c$ did not change in Mg deficient $MgB_2$ single crystals. In several studies the thermal stability of $MgB_2$ has been addressed since it is a very important factor for practical application. Fan *et al.* [22] investigated the role of thermodynamic and kinetic barriers in the decomposition process of $MgB_2$ polycrystalline samples in vacuum ( > $10^{-9}$ Torr). A significant Mg and B evaporation was observed beginning near 425 °C and 650 °C. For a given temperature, the evaporation of Mg was found to be at least two orders of magnitude higher than that of B. It was also reported that under a constant flow of argon gas at 700 °C, $MgB_2$ slowly decomposed into Mg and $MgB_4$ [23]. It seems that under our experimental conditions the kinetic barriers for Mg vaporization out of $MgB_2$ single crystals is higher than in polycrystalline samples and this leads to a higher decomposition temperature. However, the decomposition reaction of $MgB_2$ into $MgB_4$ and Mg is the same as in the vacuum annealed polycrystalline samples. On the other hand, the changes of the superconducting volume fraction of the crystals studied correlates well with increasing concentration of vacancies and they suggests a selective decrease of Mg site occupancy during annealing, leading to a phase separation into Mg vacancy-poor regions with $T_c$ of as-grown crystals and essentially non-superconducting Mg vacancy-rich regions. However one should notice that we do not have thermodynamic equilibrium in the vacuum annealing experiments. Therefore increasing the annealing time can lead to decomposition at temperatures lower than 975 °C, the temperature where we observed first signs of decomposition. The next question which should be answered is why the critical temperature remains apparently unaffected. Since each Mg atom contributes two electrons, the Mg vacancy will add two holes. It is well known that $MgB_2$ is a two-gap multi-band superconductor with a two-dimensional σ band and a three-dimensional π band and three kind of scattering mechanism are possible in this material: intraband scattering within the σ and π bands and interband scattering between the two bands [24]. The superconductivity is mainly associated with the σ band and the hole doping of the π band should not affect $T_c$. One might expect that Mg vacancies increase intraband scattering in the π band rather than in the σ band and this can be the reason why $T_c$ does not change. Indeed, recent point contact Andreev reflection measurements of polycrystalline samples [25] reveal that deliberately introducing Mg deficiency dirties the π band without introducing significant interband scattering. Earlier investigations performed on the same samples [26] also found that $T_c$ remains constant across the sample series. These observations agree with our results. However, a full explanation of the effect of Mg vacancies would require the knowledge of other relevant parameters such as the density of states, coupling and phonon frequencies.

To shed light on these issues we carried out first-principles electronic structure calculations for MgB$_2$ with Mg vacancies Mg$_{1-x}$(V$_{Mg}$)$_x$B$_2$. Our calculations are based on the density functional theory (DFT) in the generalized gradient approximation (GGA) for exchange-correlation according to Perdew and Wang [27]. We use a plane wave basis code and projected augmented wave (PAW) approach [28] provided in the VASP package [29]. We employ a 312 eV kinetic energy cutoff for the basis set, with a very soft B pseudopotential. The PAW approach provides a very good description of bulk properties of MgB$_2$, such as the lattice parameters $a$ = 3.067 Å, $c$ = 3.536 Å, in good agreement with the experimental values. We simulate the Mg vacancies within periodic boundary conditions via the supercell approach. For each defect both atomic positions and supercell lattice parameters were relaxed to the minimum energy configuration. Two supercells have been used to simulate $x$ = 0.125 and $x$ = 0.0625 molar fraction of Mg vacancies, corresponding to the Mg$_7$B$_{16}$ and Mg$_{15}$B$_{32}$ formula units. Figure 5 shows the angular momentum decomposed projected density of states (PDOS) on to B atoms for the defected systems compared with the corresponding PDOS for a stoichiometric MgB$_2$ bulk. The projection refers to a muffin tin sphere of 0.84 Å radius. On the abscissa we set the zero of the energy scale at the Fermi level.

In the upper part of the figure the three lines show the PDOS p$_{xy}$ component originating from the σ bands for the above mentioned vacancies concentrations. The vacancies do not change the shape of the density of states, they just shift the value of $E_F$ because of the electron removal. The only consequence of this shift is a modest enhancement of the σ band related density of states at $E_F$. In the lower part of the figure the PDOS p$_z$ component originating from the π bands is shown. Here, because of the flat shape of the PDOS at $E_F$ the effect of the Mg defects on the electronic structure is not evident at first sight; the π component of the DOS at $E_F$ is unchanged. Nevertheless a closer inspection shows that the onset of the π bands does shift upward by about 0.4 eV an amount twice larger than the effect on the σ bands. Overall the effect of the vacancy on the electronic structure is a shift toward higher energies of the onset of the π bands, relative to the σ bands onset. This is somewhat similar to what was observed in the case of Li substitution [30].

Further insight can be obtained by the supercell band structure dispersion shown in Fig. 6. Here we show the bands along the high symmetry direction using dots whose size is proportional to the localization of the state on a given atom. Moreover the shape of the dots is chosen to be a red square if it has a predominant $p_{xy}$ (σ) or a blue triangle if it is of $p_z$ (π) atomic angular momentum character (color online). Panel (a) and (b) refer to the *x* = 0.125 supercell and panel (c) to the fully occupied structure within a supercell. Panel (a) shows the bands with a sizeable contribution coming from the B nearest neighbor to $V_{Mg}$, while panel (b) shows those bands belonging to the farthest B to $V_{Mg}$. Because of the band folding in the supercell geometry, we find the MgB$_2$ M point to be folded at the Γ point of Fig. 6. We focus on the Γ–A high-symmetry direction: the band dispersion along the Γ–A shows correctly that the π band in the bulk (panel (c)) lies just below the σ band. We know from Fig. 5 that the most relevant effect of the defect is a downshift of $E_F$. This is confirmed by the bands drawn in panel (a) and (b). The σ band is clearly higher in energy in both panels. As for the π bands we see that there is an additional band (marked by an arrow) in panel (a) and (b). This is a state coming from the combination of $p_z$ orbital belonging to B atoms closer to the defect. Because of the delocalized nature of the π states and the high density of defects (12.5%) it keeps a bulk-like dispersion, and its contribution to the π component of the DOS is a flat term just above that of the bulk as seen in Fig. 5.

The most relevant result is the following: the π bands are higher in energy in the proximity of the defect (with an up-ward shift relative to σ bands) and nearly recover the bulk-like position away from the defect. This suggests that the holes contributed to the system by the vacancy will go to both bands. For the σ bands the holes will be delocalized while for the π bands a significant degree of localization around the defect occurs. As a consequence, the holes introduced by vacancies go only partially to σ bands, whose density of states is only slightly increased, even for heavy doping. Therefore, no decrease of $T_c$ is expected, as experimentally found. To produce an increase of $T_c$, one would need the vacancy to increase the electron-phonon coupling, an explicit calculation of the defect contribution would require a prohibitively extensive calculation.

In the present study of the influence of Mg vacancies on the structure and on the superconducting properties of MgB$_2$ single crystals show that MgB$_2$ crystals remain stable up to ~ 6 % Mg deficiency. Decreasing the Mg occupation from 0.98 to 0.92 does not decrease $T_c$, but the diamagnetic signal decreases by 60%, and the superconducting transition width slightly increases from $\Delta T_c$ = 0.55 K to 1.3 K. This indicates a decrease of the superconducting volume. Significant reduction of the superconducting volume fraction without noticeable reduction of $T_c$ for the studied single crystals can be interpreted by a selective decrease of Mg site occupancy during annealing, leading to a phase separation into Mg vacancy-poor regions with $T_c$ of as-grown crystals and essentially non-superconducting Mg vacancy-rich regions. This observation in single crystals supports the notion of percolative superconductivity proposed for polycrystalline samples [8]. Electronic structure calculations of Mg$_{1-x}$(V$_{Mg}$)$_x$B$_2$ within the supercell approach show that the PDOS is only marginally influenced and that the holes introduced by the vacancies go partially to σ bands, and partially to π states. In agreement with the present observations, $T_c$ is not expected to be much affected.

This work was supported by the Swiss National Science Foundation through the NCCR program Materials with Novel Electronic Properties (MaNEP) and the Polish Ministry of Science and Higher Education, within the research project for the years 2007-2010 (No. N N202 4132 33). F.B. acknowledges CASPUR for support by the HPC grant 2009.

**Figures and captions**

**Figure 1.** (color online) a) Diamagnetic moment *M* normalized to ZFC magnetization value at 5 K for the as-grown crystal as a function of temperature for $MgB_2$ as-grown crystal and after being annealed under vacuum (>$10^{-6}$ Torr) and temperatures between 850 and 975 °C. The measurements were performed in a field of 5 Oe upon heating from the zero-field-cooled state. The field was applied along the *c*-axis of the crystal. b) Broadening of the superconducting transition $\Delta T_c$ for high temperature vacuum annealed $MgB_2$ single crystals.

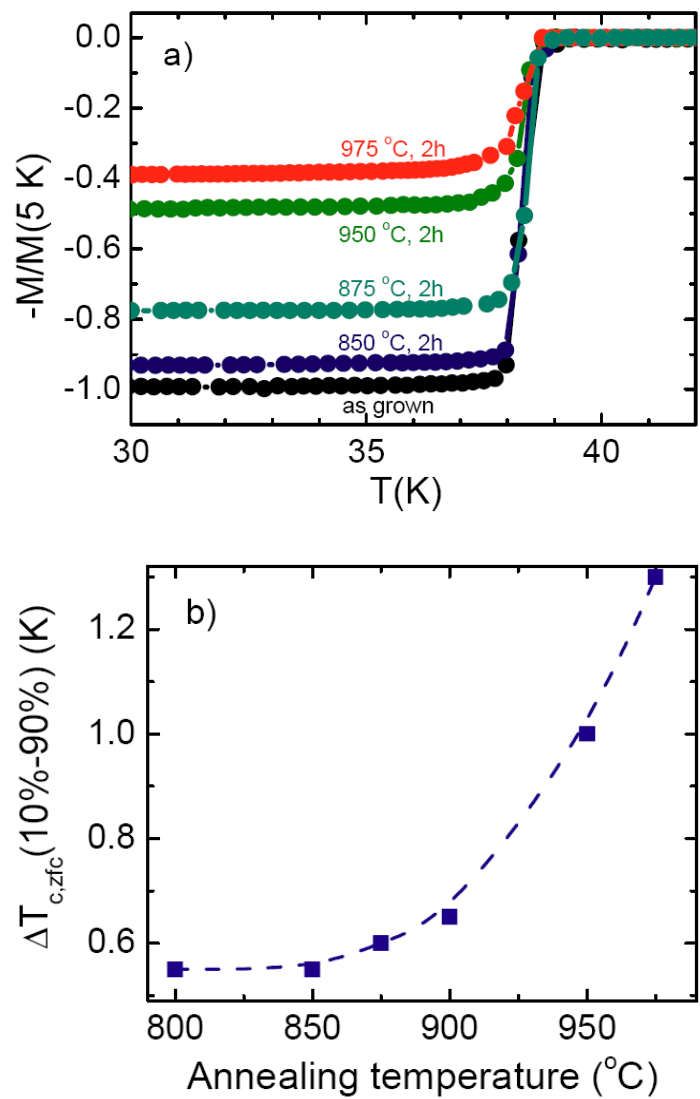

**Figure 2.** (color online) The reconstructed hk0 reciprocal space sections of the MgB$_2$ samples vacuum annealed at 850 °C (a), (b), and 950 °C (c). Rotation pictures (step 1°) of the samples annealed at 975 °C (d) and 1000 °C (e). The reconstructed powder pattern (from 2D to 1D) of the sample annealed at 1000 °C (f), indicating the presence of MgB$_4$.

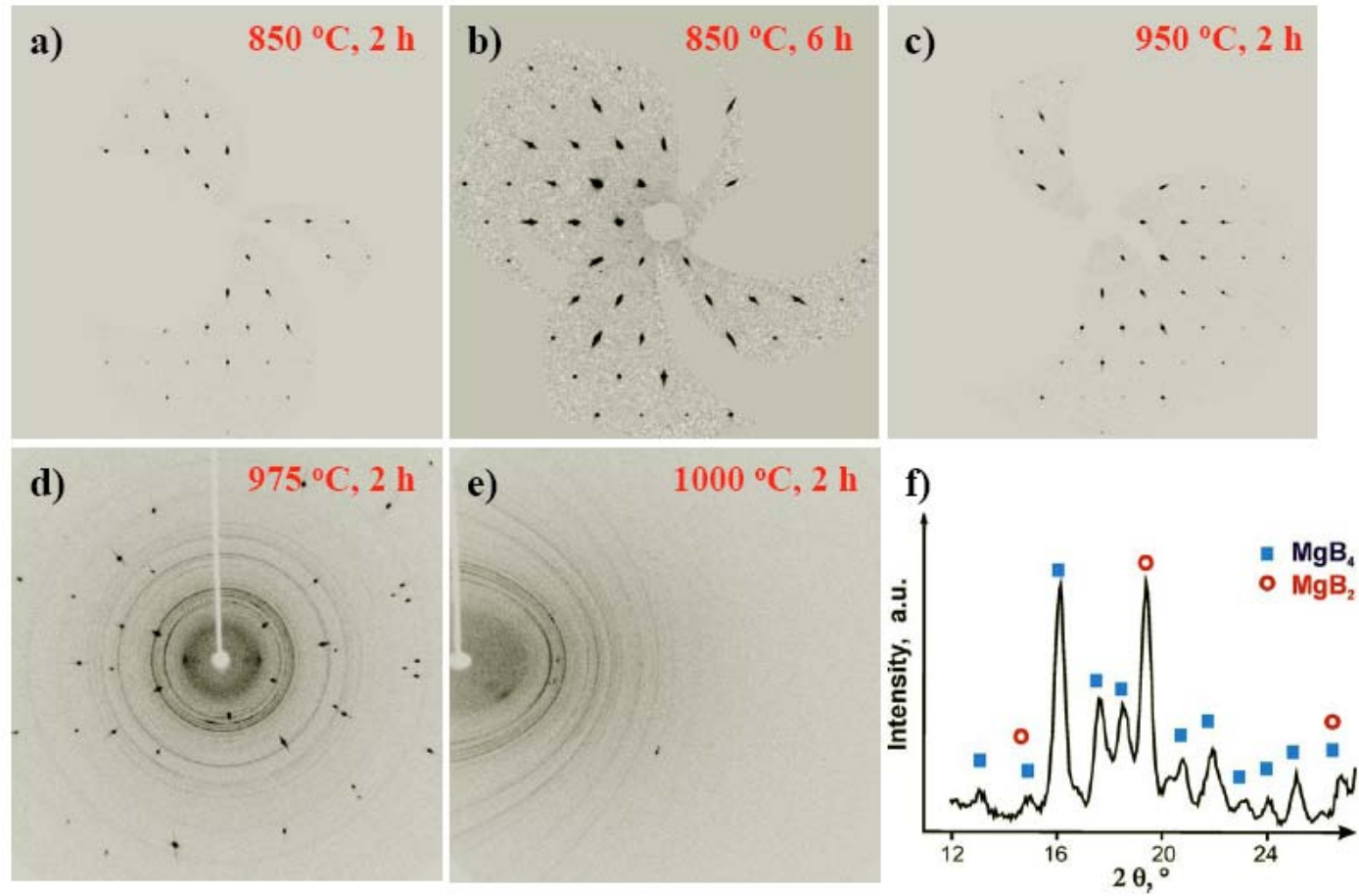

**Figure 3.** The lattice parameters *a* and *c* as a function of Mg deficiency in $Mg_{1-x}B_2$ crystals. Open symbols are taken from ref. [21].

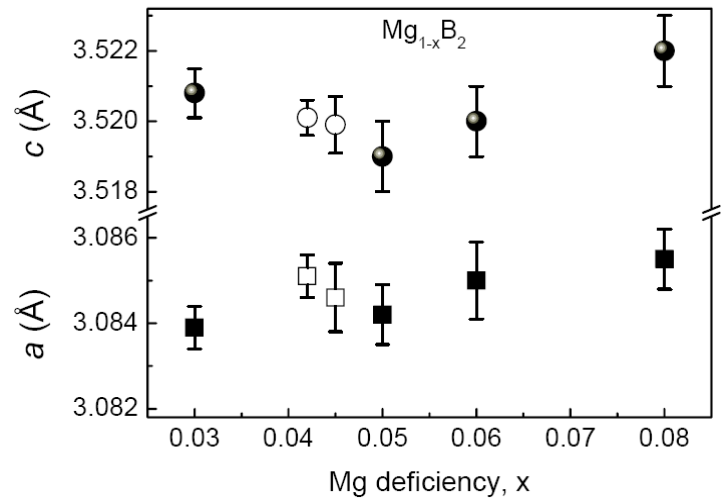

**Figure 4.** (color online) Mg occupation and relative diamagnetic signal in $MgB_2$ single crystals as a function of annealing temperature in vacuum.

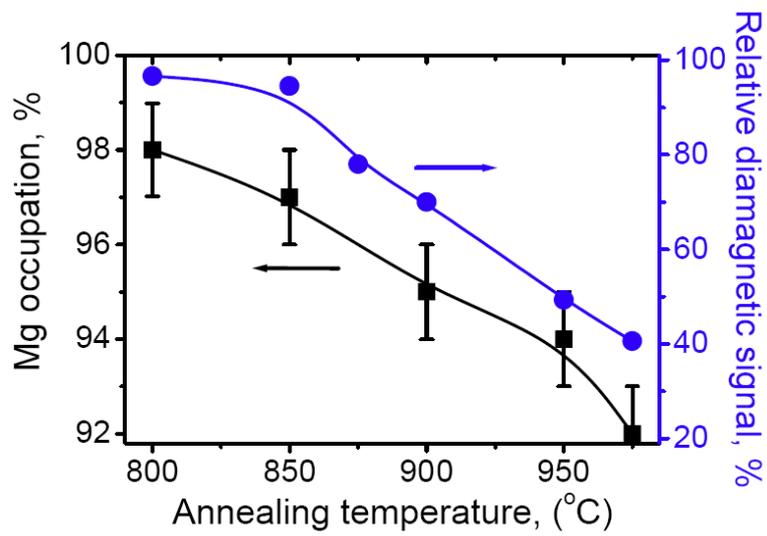

**Figure 5.** (color online) Density of states projected on the B atoms. Solid, dot-dashed and dotted lines show the PDOS for $V_{Mg}$ fraction *x* of 0, 1/16, and 1/8. The red and blue lines show the σ and π components of the bonding.

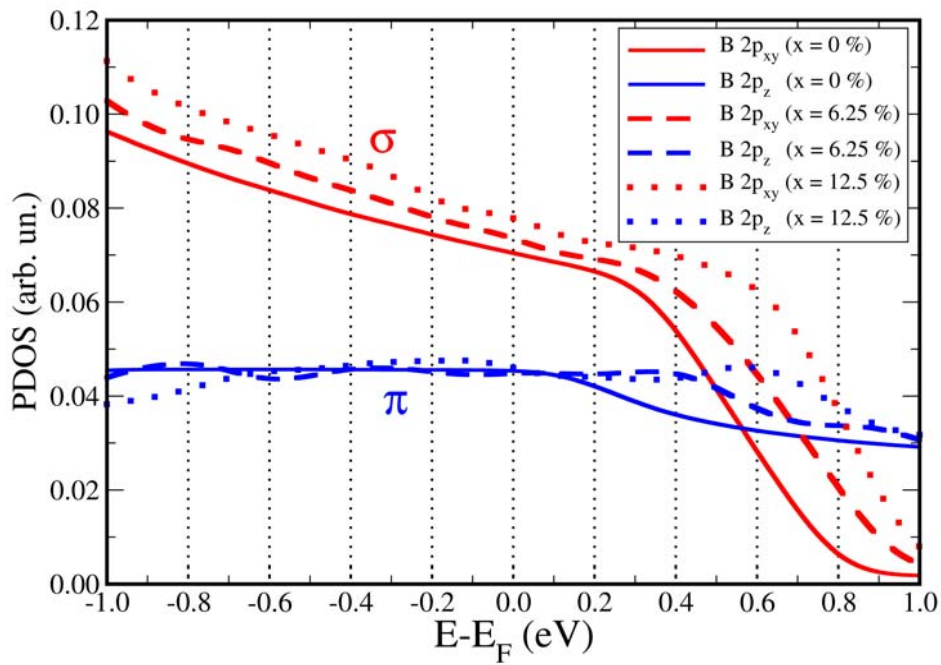

**Figure 6.** (color online) Supercell band structure along the high symmetry directions for $Mg_8B_{16}$ and $Mg_7B_{16}$ : $V_{Mg}$ systems inside a 222 supercell. Dots size is proportional to the PDOS on the B nearest neighbor to $V_{Mg}$ (panel (a)), B furthers to $V_{Mg}$ (panel (b)). Panel (c) shows the stoichiometric bulk band structure for reference. Red squares (blue triangles) refers to σ (π) band character. The arrows show the band arising from the interaction of the B $p_z$ orbitals with $V_{Mg}$.

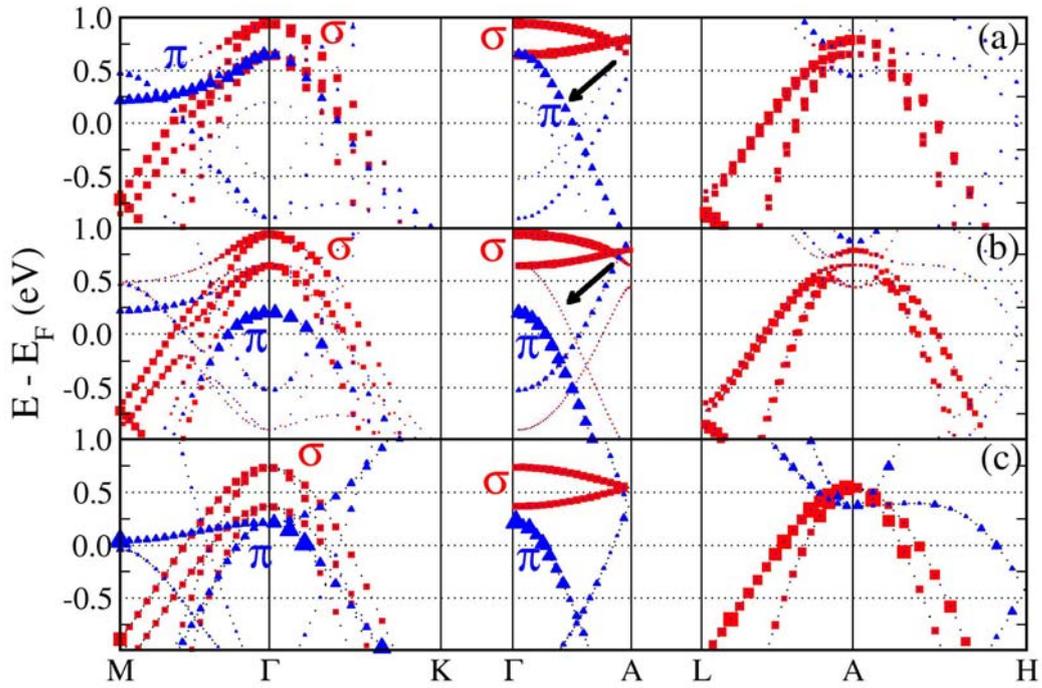

**Table 1.** Details of the structure refinement for a $Mg_{1-x}B_2$ crystals after successive steps of vacuum annealing. The diffraction study was performed at 295(2) K using Mo $K_\alpha$ radiation with $\lambda$ = 0.71073 Å. The lattice is hexagonal, *P6/mmm* space group with *Z* = 1. The absorption correction was done analytically. A full-matrix least-squares method was employed to optimize $F^2$.

| Sample | AN307/9, 850 °C | AN307/9, 900 °C | AN307/9, 950 °C | AN307/9, 975 °C |
|---|---|---|---|---|
| Mg site occupation | 0.97 | 0.95 | 0.94 | 0.92 |
| $T_{c,onset}$, K | 38.7 | 38.7 | 38.7 | 38.7 |
| Formula weight | 45.20 | 44.71 | 44.47 | 43.99 |
| Unit cell dimensions, Å | $a$ =3.0839(5) | $a$ =3.0842(7) | $a$ =3.0850(9) | $a$ =3.0855(7) |
|  | $c$ =3.5208(7) | $c$ =3.5190(10) | $c$ =3.5200(10) | $c$ =3.5220(10) |
| Volume, Å$^3$ | 28.998(9) | 28.989(12) | 29.012(15) | 29.038(12) |
| Calculated density, g/cm$^3$ | 2.588 | 2.561 | 2.545 | 2.515 |
| Absorption coefficient, mm$^{-1}$ | 0.598 | 0.586 | 0.580 | 0.568 |
| F(000) | 22 | 21 | 21 | 21 |
| Crystal size, mm | 0.77 x 0.50 x 0.09 | 0.77 x 0.50 x 0.09 | 0.77 x 0.50x 0.09 | 0.77 x 0.50 x 0.09 |
| Θ range for data collection, deg | 5.79 to 36.47 | 5.80 to 36.06 | 5.79 to 37.56 | 5.79 to 37.55 |
| Limiting indices | $-4 \leq h \leq 4$, $-4 \leq k \leq 4$, $-5 \leq l \leq 3$ | $-4 \leq h \leq 5$, $-3 \leq k \leq 5$, $-5 \leq l \leq 4$ | $-4 \leq h \leq 5$, $-5 \leq k \leq 4$, $-2 \leq l \leq 6$ | $-5 \leq h \leq 5$, $-4 \leq k \leq 3$, $-3 \leq l \leq 5$ |
| Reflections collected/unique | 175/39, $R_{int}$ = 0.0164 | 189/46, $R_{int}$ = 0.0245 | 332/48, $R_{int}$ = 0.0217 | 246/48, $R_{int}$ = 0.0380 |
| Max. and min. transmission | 0.945 and 0.780 | 0.951 and 0.808 | 0.947 and 0.722 | 0.957 and 0.767 |
| Data /restraints/parameters | 39/0/6 | 46/0/6 | 48/0/6 | 48/0/6 |
| Goodness-of-fit on $F^2$ | 1.275 | 1.192 | 1.377 | 1.345 |
| Final *R* indices [$I>2\Omega(I)$] | $R_1$ = 0.0357 | $R_1$ = 0.0300 | $R_1$ = 0.0217 | $R_1$ = 0.0197 |
|  | $wR_2$ = 0.0883 | $wR_2$ = 0.0681 | $wR_2$ = 0.0625 | $wR_2$ = 0.0600 |
| *R* indices (all data) | $R_1$ = 0.0367 | $R_1$ = 0.0313 | $R_1$ = 0.0224 | $R_1$ = 0.0243 |
|  | $wR_2$ = 0.0885 | $wR_2$ = 0.0683 | $wR_2$ = 0.0628 | $wR_2$ = 0.0603 |
| $\Delta\rho_{max}$ and $\Delta\rho_{min}$, e/Å$^3$ | 0.266 and -0.587 | 0.286 and -0.368 | 0.257 and -0.410 | 0.449 and -0.251 |